# High Accuracy and Low Complexity Frequency Offset Estimation Based on All-Phase FFT for M-QAM Coherent Optical Systems

Qian Li[1], Aiying Yang[1,*], Peng Guo[1], Xiangjun Xin[2]

*1 Key Laboratory of Photonics Information Technology, Ministry of Industry and Information Technology,
School of Optics and Photonics, Beijing Institute of Technology, Beijing 100081, China*
*2 School of Electronic Engineering, Beijing University of Posts and Telecommunications, Beijing 100876, China*
*\* yangaiying@bit.edu.cn*

**Abstract:** A low complexity frequency offset estimation algorithm based on all-phase FFT for M-QAM is proposed. Compared with two-stage algorithms such as FFT+CZT and FFT+ZoomFFT, our algorithm can lower computational complexity by 73% and 30% respectively, without loss of the estimation accuracy.

## 1. Introduction

Digital signal processing (DSP) algorithms can compensate for various types of impairments in coherent optical communication systems [1], thus greatly improving the sensitivity of coherent optical receiver, and can achieve higher spectral efficiency and bit rate transmission combined with high-order modulation format [2], e.g., M-ary quadrature amplitude modulation (M-QAM). The frequency offset estimation (FOE) is a key DSP algorithm to compensate for phase rotation caused by the frequency mismatch between the transmitter and the local oscillator (LO) Lasers. The widely used FOE algorithms can be divided into two categories. The first type relies on differential phase (diff-FOE) [3]. It has relatively low computational complexity, but its accuracy is poor and not suitable for high-order modulation formats such as 64QAM. The other type is based on fast Fourier transform (FFT-FOE) [4]. This algorithm requires a large number of symbols to reach sufficient spectral resolution and results in high complexity. To reduce the complexity, two-stage FOE algorithms, such as CZT or ZoomFFT, are proposed [5, 6]. In these algorithms, the first step of coarse FOE is carried by operating FFT on a few symbols. Then, the CZT or ZoomFFT are used in second step to accurately estimate the frequency offset. However, the computational complexity of CZT or ZoomFFT is still high.

In this letter, we propose an all-phase FFT (apFFT) based FOE algorithm. The apFFT has the advantages of spectrum leakage suppression and high phase estimation accuracy [7]. Our proposed algorithm uses the phase information of apFFT spectrum to correct the measured frequency offset and improves the accuracy of FOE. Simulation results on 28GBaud optical 16/64-QAM transmission systems show that the proposed apFFT based FOE algorithm can achieve high estimation accuracy with lower complexity.

## 2. Principle of apFFT Based Algorithm

The 4th power of $n$th received M-QAM symbol of coherent signal can be expressed as [4]

$$S_n^4 = A\exp[4j(2\pi f_d nT_s + \varphi_{l,n})] + e_n \tag{1}$$

where A is a non-zero constant amplitude (e.g. for 16QAM, $A = -68$); $f_d$ is the frequency offset between the transmitter and the LO; $T_s$ represents the symbol duration; $\varphi_{l,n}$ is the Wiener process phase noise and $e_n$ is a zero-mean process that can be considered as disturbing noise. If the effect of phase noise $\varphi_{l,n}$ on signal spectrum is ignored, it can be regarded as a single frequency complex exponential signal, then we can perform apFFT on the signal.

Fig. 1 shows the schematic diagram of apFFT [7]. A convolution window $w_c$ of length 2N−1 is used to weight the 2N−1 signals before and after the central sample point $x(0)$, then the weighted signals with an interval of N are added to form N signals, and then FFT is performed on these N signals to obtain the apFFT spectrum of the original 2N−1 signals. The apFFT of $S_n^4$, denoted by $Y_{ap}(k)$, can be expressed as

$$Y_{ap}(k) = \left[\frac{A}{N}\frac{\sin[\pi(4f_d NT_s - k)]}{\sin[\pi(4f_d T_s - \frac{k}{N})]}\right]^2 e^{j\theta_0} \tag{2}$$

where $\theta_0$ denotes the phase of the central sample point $S_0^4$. From (2), a coarse estimate of frequency can be obtained based on the position of apFFT spectrum peak, expressed as $f_{\text{coarse}} = \frac{\hat{k}}{NT_s}$, which is the same as FFT-FOE, but its estimation accuracy is also limited by the picket-fence effect of FFT.

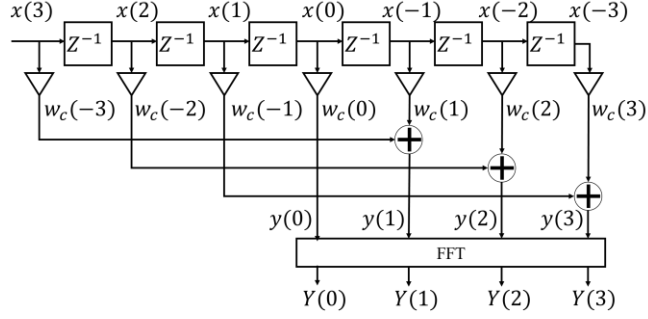

Fig. 1. Schematic diagram of apFFT spectral analysis (in the case of N=4)

Next, FOE is carried out for the next block of signals with a delay of $NT_s$ seconds (2N−1 signals before and after the central sample point $S_N^4$), and the phase at the peak of apFFT spectrum is $\theta_N$, which can be expressed as

$$\theta_N = \theta_0 + 2\pi \cdot 4f_d NT_s \qquad (3)$$

The frequency offset $4f_d$ can be divided into integer and fractional parts according to apFFT spectral resolution $\frac{1}{NT_s}$, and expressed as

$$4f_d = (k_d + \delta)\frac{1}{NT_s}, \quad -0.5 < \delta < 0.5 \qquad (4)$$

Based on (3) and (4), the estimated value for $\delta$ can be described as $\delta = \frac{\theta_N - \theta_0}{2\pi}$. Finally, the frequency offset can be obtained as

$$\hat{f}_d = \frac{1}{4}\left(f_{\text{coarse}} + \frac{\delta}{NT_s}\right) \qquad (5)$$

## 3. Proof Results of apFFT Based Algorithm

The apFFT based algorithm is verified by a 28Gbaud back to back 16/64-QAM coherent optical communication system build with commerical software VPI Transmission Maker. The laser linewidth of transmitter and LO are set to be 100 kHz. At the coherent receiver side, the FOE is performed with the proposed algorithm. For comparison, the two-stage algorithms, CZT/ZoomFFT are also performed. For the two-stage algorithms, the sample point size of FFT in the first stage, $N_1$, is 512 (1024) for 16-QAM (64-QAM), and the sample point size of CZT/ZoomFFT in the second stage, $N_2$, is 256 (512) for 16-QAM (64-QAM), which is consistent with the literature [5, 6]. The sample point size of the proposed apFFT based algorithm is equal to $N_1$.

Figs. 2(a) and (b) depict the normalized mean square error (MSE, defined as $E[|(\hat{f}_d - f_d)T_s|^2]$) of different FOE algorithms versus the frequency offset for 16/64-QAM, respectively. The proposed algorithm and the two-stage algorithms have the same estimation range of [-symbol rate/8, +symbol rate/8], i.e. [-3.5 GHz, +3.5 GHz] for 28 Gbaud 16/64-QAM signals. Due to the limited spectral resolution, the single-step FFT-FOE algorithm shows a large fluctuation, while the proposed algorithm and the two-stage algorithms are relatively stable in the whole range of frequency offset, and the normalized MSE reach $3.0 \times 10^{-9}$ for 16QAM and $1.6 \times 10^{-9}$ for 64QAM.

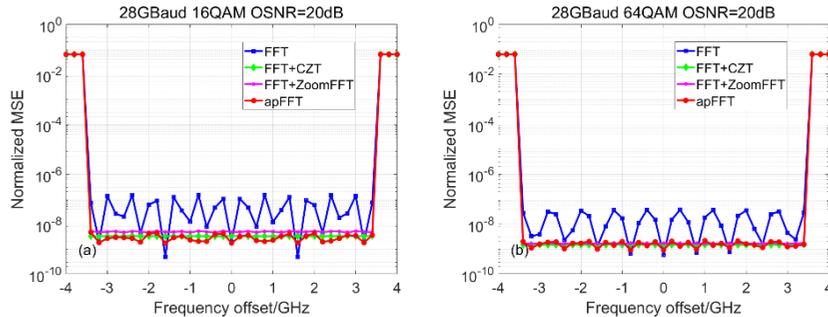

Figure. 2. Normalized MSE versus frequency offset $f_d$ for (a) 16QAM and (b) 64QAM (1000 simulations for each $f_d$)

Figs. 3(a) and (b) show the normalized MSE of different FOE algorithms as a function of OSNR. For different OSNR, the frequency offset is set from [-3.5 GHz, +3.5 GHz] with an interval of 200MHz. The simulation times for each frequency offset is 1000, so the normalized MSE under each OSNR is obtained by 36000 simulations. The results show that the OSNR threshold for our proposed algorithm and the two-stage algorithms is the same. When

OSNR is greater than the threshold, our proposed algorithm can achieve the same estimation accuracy as the two-stage algorithms.

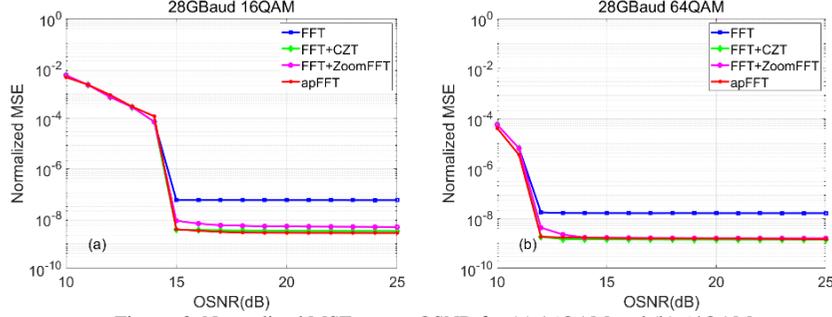

Figure. 3. Normalized MSE versus OSNR for (a) 16QAM and (b) 64QAM

According to the simulation results, the algorithm we proposed can achieve the same performance as the two-stage algorithms. Besides, the computational complexity of the proposed algorithm is lower, which will be discussed in the next.

**4. Complexity Analysis**

The number of real multiplications (denoted as MUL) for the three FOE algorithms are discussed to compare their complexity. Firstly, the 4$^{th}$ power and FFT of $N_1$ signals requires $6N_1 + 2N_1 log_2(N_1)$ MUL. Then, the MUL required by $N_2$ point CZT and ZoomFFT can be expressed as (6) and (7) according to [5, 6], respectively.

$$MUL_{CZT} = \left(\frac{106}{9} + \frac{4}{3}\log_2 L + 2\log_2 N_2\right) \times L + 14 + (-1)^{\log_2 L} - 8N_2 \quad (6)$$

$$MUL_{ZoomFFT} = 2N_2 log_2 N_2 + 8N_1 \quad (7)$$

where $L = 2\hat{}(nextpow2(N_1 + N_2 - 1))$. In comparison, our proposed algorithm only needs $2 \times (2N_1 - 1)$ MUL to window the signal in addition to the FFT and 4$^{th}$ power operations. Table 1 show the complexity of the three FOE algorithms under different parameter settings. Compared with CZT and ZoomFFT, the complexity of our proposed FOE algorithm is reduced by about 73% and 30% for 16QAM and 64QAM signal respectively.

Table 1. Multiplication Complexity

| $N_1$, $N_2$ | $MUL_{CZT}$ | $MUL_{ZoomFFT}$ | $MUL_{apFFT}$ |
|---|---|---|---|
| $N_1$=512, $N_2$=256 | 52353 | 20480 | 14334 |
| $N_1$=1024, $N_2$=512 | 113563 | 44032 | 30718 |

**5. Conclusions**

We proposed a low complexity FOE algorithm based on apFFT, which achieves the same estimation accuracy as the two-stage FOE algorithms based on CZT/ZoomFFT. Furthermore, its complexity is reduced by nearly 73% and 30% compared with the CZT and ZoomFFT based algorithms, respectively.

**Acknowledgement**

This work was supported by the National Natural Science Foundation of China under Grant 61427813, the Open Fund of IPOC (BUPT) under Grant IPOC2018B003.